\documentclass[1p,11pt]{elsarticle}
\begin{document}
\begin{frontmatter}

\title {Lorentz-invariant and Lorentz-non-invariant aspects \\
of a scalar tachyon field Lagrangian \\
and the scalar tachyon Feynman propagator}                         
%\title {Tachyon Feynman propagator in a tachyon scalar field model  \\
%         conserving causality       \\
%with a covariant Lorentz symmetry breaking}
%========================================================================%

\author[itep]{Vassili F. Perepelitsa}
\address[itep]{Institute for Theoretical and Experimental Physics, Moscow, Russia\\}
%Present address: V.F.~Perepelitsa, CERN, CH-1211 Gen\`{e}ve 23, Switzerland\\}
%Phone number 41 22 7673114\\}
%This article is registered under preprint number: /gen-ph/1605.03425}
%e-mail: vassili.perepelitsa@cern.ch
%========================================================================%

\begin{abstract}
A consistent theory of faster-than-light particles
(tachyons) can be built replacing the standard Lorentz-invariant approach
to quantum field theory of tachyons by a Lorentz-covariant one,
invoking a concept of the preferred reference frame. This is a mandatory
condition imposed by causality conservation.
In this article some features of a Lorentz-violating (but Lorentz-covariant) 
Lagrangian of a scalar tachyon field are considered. 
It is shown that the equation of motion and Feynman propagator
resulting from it are Lorentz-invariant, while the Lorentz symmetry 
of the suggested tachyon field model can be defined as covariantly broken.
%=========================================================================%
\end{abstract}

\begin{keyword}
tachyons \sep causality \sep tachyon vacuum \sep Lorentz-non-invariance \sep Feynman propagator\\
%arXiv: 1605.03425
\end{keyword}
\end{frontmatter}
\pagestyle{plain}
%=========================================================================%

\section{Introduction}
\setcounter{equation}{0}
\renewcommand{\theequation}{1.\arabic{equation}}
The generally accepted opinion existing inside the physics community is
that faster-than-light signals and particles (tachyons, \cite{bds,fein}) 
would lead inevitably to causality 
violations~\footnote{Often causality violation is related 
intrinsically to the mere possibility of existence of tachyons
by pointing out that the positive time interval between events connected by a
space-like world line can be converted to a negative time interval by a
suitable Lorentz transformation. Such a change of the event time order can
indeed take place in the case of tachyon signals, but {\em it does not mean yet
causality violation}. The causality violation appears only when a causal
loop can be constructed, i.e. a sending by an observer a signal to its own past,
see \cite{ttheor}.}. Another serious problem related to
the tachyons is the presumable instability of the tachyon vacuum. Sometimes
also the violation of the unitarity by tachyons interacting with ordinary 
particles is declared \cite{unita,unita2}.

Meanwhile it is known since the 1970's that tachyons do not violate 
causality if one postulates the existence of a preferred reference frame in
which the propagation of free tachyons is ordered by retarded causality,
see {\em e.g.} \cite{sigal,pvmich}. 
%(and later suggestions \cite{rembl,radzik}).  

With similar ideas it has been shown recently \cite{ttheor} that the tachyon 
hypothesis, if one treats tachyons properly, does not lead to the appearance 
of causal paradoxes, while the causality principle has to be redefined 
as a requirement of the absence of causal loops, i.e. the
impossibility of a transfer of information to the past of an observer. The 
resolution of causal paradoxes comes, indeed, from a trivial observation that
fast tachyons can probe cosmological distances. This leads to a conclusion that
the completely correct description of tachyon behaviour can be made within
general relativity only, while the Lorentz symmetry has to be considered as
an approximate one when applied to tachyons.  This makes the use of 
tachyons for the construction of the causal loops impossible (the reasons for 
this are formulated in the next paragraph). It has been shown in \cite{ttheor}
that the correct approach to the tachyon theory can be achieved only within the
postulate of a tight association of the tachyon preferred reference frame with 
the comoving frame of relativistic cosmology (see the definition of the 
latter, for example, in \cite{ll1}). This is the absolute rest frame \cite{pdg}
in which our universe is embedded. In particular, the distribution of matter 
in the universe is isotropic in this frame only, the same is true for 
the relic black body radiation. 
The causality protection formula, valid in all inertial frames, was formulated
in \cite{ttheor} as follows:
\begin{equation}
       Pu \geq 0,
\end{equation}
where $P$ is a 4-momentum of particles transferring a signal and $u$ is a
4-velocity of the preferred reference frame with respect to (any particular)
inertial observer. It is a boundary condition which should be imposed on
solutions of any tachyon equation of motion.

The next step, the introduction of the concept of the preferred reference frame 
into the Minkowski space, can be considered as an action approximating that
space to the real world space-time. It does not destroy the mathematical
perfectness of the Lorentz group since the derivation of the Lorentz 
transformations is based on the requirement of the invariance of the interval 
between world points when passing from one inertial frame to another, 
and the presence or the absence of the preferred reference frame 
among the frames under consideration does not affect the derivation 
to any extent. In view of this one can retain the Lorentz group (as well as 
the Poincar\'{e} group) when treating tachyons. Together with this the 
introduction of the preferred reference frame into the tachyon theory removes 
the problem of the instability of the tachyon vacuum (see \cite{ttheor}
and sect.~2 in this article). 
The space of the preferred frame turns out to be 
spanned by the continuous background of free, zero-energy on-mass-shell tachyons
propagating isotropically, i.e. the eigenvalues of the tachyon Hamiltonian 
are restricted from below, in this frame, by zero value. This excludes
the possibility of the construction of the causal loops using tachyons 
since for such a construction negative energy tachyons (propagating 
backward in time) are necessary. In the frames moving with respect to the
preferred one negative energy tachyons can appear due to Lorentz boosts,
but causal loops cannot, since the presence or the absence of the causal 
loops is an invariant property of the relativistic theory describing tachyons.
For example, the energy boundaries of the tachyon vacuum, which has to be 
defined in tachyon quantum field theories by the tachyon vacuum gauge
\begin{equation}
     Pu = 0,
\end{equation}
in the frames moving with respect to the preferred one 
%with the velocity $u = (u^0,{\bf u})$ 
are given by expressions
\begin{equation}
E_0^+ = \frac{\mu |{\bf u}|}{\sqrt{1 - {\bf u}^2}}
\end{equation}
for the direction coinciding with the preferred frame velocity ${\bf u}$ and by 
\begin{equation}
E_0^- = -\frac{\mu |{\bf u}|}{\sqrt{1 - {\bf u}^2}}.
\end{equation}  
for the opposite direction.

Simultaneously it turns out that in any reaction in which tachyons participate
asymptotic ``in" and ``out" tachyonic Fock spaces are unitarily equivalent,
which solves the unitarity problem.

It was argued in \cite{ttheor} that a realistic 
model of a tachyon theory should be built upon the infinite-dimensional
unitary irreducible representations of the Poincar\'{e} group
(so called ``infinite spin" tachyons). Within the conjecture that elementary
particles are realizations of the unitary irreducible representations of the
Poincar\'{e} group the only alternative to the infinite spin tachyon models 
would be a scalar tachyon model. However this model cannot represent tachyons 
at a fundamental level since it possesses several diseases; in particular, 
such a model would lead to the instability of photons via their decay to
tachyon-antitachyon pairs \cite{ttheor} (note that decays of photons to the 
infinite spin tachyon-antitachyon pairs are forbidden by the angular momentum 
conservation combined with kinematic restrictions imposed on this process). 

Nevertheless, a scalar tachyon field model has to be considered first for 
the following reasons. Being simpler than a theory based on the 
infinite-dimensional wave equation, it can serve as a toy model in which 
several new theoretical ideas can be tested, namely, the realization of the 
postulate of a preferred reference frame as applied to quantum field theory
in a covariant way, the construction of a stable tachyon vacuum with a
confinement of acausal tachyon modes, the Lorentz-invariance of the tachyon 
Feynman propagator, necessary to restrict Lorentz-violating effects to the 
free-tachyon sector only, etc. Moreover, all individual components of the
infinite-dimensional wave equation must satisfy the Klein-Gordon equation.   

Therefore the scalar tachyon models were considered in ref. \cite{ttheor}.
They are based on Lorentz-covariant scalar tachyon Lagrangians 
with a covariant breaking of the Lorentz symmetry, so the 
Lorentz invariance violation appears to 
%be restricted to the tachyon sector only, affecting 
be affecting the asymptotic tachyon states only, 
while leaving the sector of ordinary particles within the Standard Model 
untouched, at least up to possible small radiative corrections. 
For example, the Hermitian tachyon field operator with the causal 
$\Theta$-function accounting for the boundary condition (1.1), $\Theta(ku)$, 
reads as follows:
\begin{equation}
\Phi(x) = \frac{1}{\sqrt{(2\pi)^3}} 
\int{d^4k~\Big{[}a(k)\exp{(-ikx)} + 
a^+(k)\exp{(ikx)}\Big{]}~\delta(k^2+\mu^2)~\Theta(ku)},
\end{equation}
where $k$ is a tachyon four-momentum, $a(k), a^+(k)$ are annihilation and
creation operators with bosonic commutation rules, annihilating or creating 
tachyonic states with 4-momentum $k$, and $\mu$ is a tachyon mass parameter.
As can be seen, the expression (1.5) is explicitly 
Lorentz-covariant. This covariance includes the invariant meaning of the 
creation and annihilation operators defined in the preferred frame;
thus, for example, an annihilation operator $a(k)$ remains an
annihilation operator $a(k^\prime)$ in the boosted frame, even if the zero
component of $k^\prime$ may become negative. This is because the one-sheeted
tachyon mass-shell hyperboloid is divided by the covariant boundary
$(ku) = 0$ into two parts separated in an invariant way. 
This results, in particular, in a possibility of the standard operator
definition of the invariant vacuum state $|0>$ via the annihilation operators
$a(k)$, $a(k)|0> = 0$ for all $k$ such that $|{\bf k}| > \mu$, because the 
vacuum state energy and, as a consequence, the field Hamiltonian turn out 
to be bounded from below in any reference frame,
see \cite{ttheor} and formulae (\ref{eq:210}), (\ref{eq:211}) below. 

%When calculating the tachyon production probabilities and cross-sections
%the confining $\Theta$ functions will accompany the production amplitudes as
%factors restricting the reaction phase space, so the expressions for
%these probabilities can be displayed as follows:
%\begin{equation}
%W = \int| M |^2 d\tau \prod_{i} \Theta(k_i u),
%\end{equation}
%where M is a matrix element of the reaction (which has to be representable in a
%Lorentz-invariant form), $d\tau$ is a reaction phase space element, and the
%product of $\Theta$ functions includes all tachyonic asymptotic states
%(having 4-momenta $k_i$) participating in the reaction \cite{pvmich}.

In paper \cite{causal} a formal way of introducing the causal 
$\Theta$-function into the tachyon field operator (1.5) has been suggested. 
In this article we continue the consideration of Lorentz-invariant and 
Lorentz-non-invariant properties of the tachyon field models mentioned above,
including an important element of the models such as the Lorentz-invariance 
of the Feynman propagator of the tachyon scalar fields.

This note is organized as follows. In Section~2 we suggest a
Lorentz-non-invariant, but Lorentz-covariant modification of the scalar
tachyon Lagrangian which leads to a tachyon Hamiltonian possessing
Lorentz-non-invariant boundaries of the tachyon vacuum. The modified
Lagrangian leaves however the tachyon equation of motion unchanged which
leads to the Lorentz invariance of the tachyon Feynman propagator considered
in Section~3. Representations of this propagator in the configuration
space are given in Section~4. In Section~5 some Lorentz-invariant and 
Lorentz-non-invariant two-point functions of scalar tachyon fields are 
presented. Sections~6 and~7 contain the note summary and conclusion.

In formulae used in this article the velocity of light $c$ and the Planck
constant $\hbar$ are taken to be equal to 1.

%\section{Stability of the tachyon vacuum and 
\section{Proposed Lorentz-non-invariance of the tachyon Lagrangian}
\setcounter{equation}{0}
\renewcommand{\theequation}{2.\arabic{equation}}
We start with a Lorentz-invariant Lagrangian of a free scalar tachyon field
\begin{equation}
L = \frac{1}{2} \int d^3 {\bf x} \Big{[}\dot\Phi^2(x) -
\Big{(}\nabla\Phi(x)\Big{)}^2  + \mu^2 \Phi^2(x) \Big{]}
\end{equation}
resulting in the (Klein-Gordon) equation of motion 
\begin{equation}
\Big{(}\frac{\partial ^2}{\partial t^2} - \partial _i \partial ^i - \mu^2\Big{)} \Phi(x) = 0~, ~~~~~~~~i = 1,2,3
\end{equation}
and in the Hamiltonian
\begin{equation}
H = \frac{1}{2} \int d^3 {\bf x} \Big{[}\dot\Phi^2(x) +
\Big{(}\nabla\Phi(x) \Big{)}^2 - \mu^2 \Phi^2(x)\Big{]}.  
\end{equation}
A standard approach to finding the minimum of the Hamiltonian (2.3), 
which could present the field ground state $\Phi_0$, i.e. the vacuum 
(requiring $\delta H = 0$), is reduced to an analysis of its potential 
term. This assumes, implicitly, that the search for the ground state of the
Hamiltonian is replaced by looking for its minimum under restrictions
conditioned by the Lorentz-invariant pair of the vacuum ``initial" conditions:
\begin{equation}
\Phi_0 = const~in~time,
\end{equation}
\begin{equation}
\Phi_0 = uniform~(const)~in~space.
\end{equation}
In the case of ordinary particles, with a positive $m^2$ in the corresponding 
Hamiltonian, the exercise of ground state finding under these conditions 
succeeds at $\Phi = 0$, but the potential term of the Hamiltonian (2.3) has 
a maximum at $\Phi = 0$ instead of the necessary minimum. This is interpreted 
as an impossibility (instability) of the tachyon vacuum.

However, the hypothesis of the faster-than-light particles requires the
consideration of tachyons under a postulate of a preferred reference frame
which is necessary for the causal ordering of the signals propagating over
the spacelike intervals. This requirement must be respected by the procedure
of the tachyon ground state finding also. Therefore the Lorentz-invariant pair
of the initial conditions (2.4), (2.5) must be replaced by a single,
Lorentz-non-invariant one:
\begin{equation}
\Phi_0 = const~in~time
\end{equation}
which separates, obviously, the preferred reference frame, while the condition
(2.5) should be avoided, thus excluding spatially uniform fields from 
the consideration . Then the Hamiltonian which has to be analysed 
in the search for the ground state will contain,
together with the potential term, a gradient energy term which compensates 
the former. The equation $\delta H = 0$ becomes equivalent to the equation
$\delta L = 0$, i.e. to the equation of motion. In our case it is the 
Klein-Gordon equation (2.2) which has, in general, 
solutions in the form of plane waves
$\exp{ \pm i(Et - {\bf kx})}$ with the dispersion relation
\begin{equation}
E \equiv k^0 = \sqrt{\bf{k}^2 - \mu^2},
\end{equation}
under the prescription \cite{bds,fein} 
\begin{equation}
|\bf{k}| \geq \mu 
\label{eq:8}
\end{equation}
(note, this prescription is Lorentz-invariant \footnote{To prove this statement
let us consider the inequality $|{\bf{k}}| \geq \mu$, valid in some inertial 
frame, in a boosted frame. In order to simplify the proof  we take the boost 
direction opposite to the tachyon momentum ${\bf{k}}$: this is a critical case 
for the validity of the inequality; then it transforms to 
$(|{\bf{k}}| - E |{\bf{u}}|)/\sqrt{1-{\bf{u}}^2} \geq \mu$.
This can be rewritten as
\begin{equation}
\frac{|{\bf{v}}| - |{\bf{u}}|}{\sqrt{{\bf{v}}^2 -1}\sqrt{1-{\bf{u}}^2}} \geq 1,
\label{eq:9}
\end{equation}
where $\bf{v}$  is a tachyon velocity, $\bf{v}=\bf{k}/E$.
Consider now the function of modules of 3-dimensional vectors
${\bf{u}}$ and ${\bf{v}}$, $f(u,v) = (v-u)/(\sqrt{v^2-1}\sqrt{1-u^2})$.
This function reaches its minimal value of 1 at $uv = 1$. 
%Its partial derivatives $f^{'}_u, f^{'}_v$ have the common solutions of 
%equations $f^{'}_u = 0$, $f^{'}_v = 0$ at $uv = 1$, which minimizes $f(u,v)$ 
%to 1. We have to consider also the four extreme cases: 
%$|{\bf{u}}| \rightarrow 0$ and $|{\bf{u}}| \rightarrow 1$ at arbitrary 
%${\bf{}v}$, and $|{\bf{v}}| \rightarrow 1$ and $|{\bf{v}}| \rightarrow \infty$
%at arbitrary ${\bf{u}}$. The direct check shows that in all four cases 
%the expression (2.9) is valid. 
This means that the expression (\ref{eq:9}) holds always,
i.e. the condition $|\bf{k}| \geq \mu$ is Lorentz-invariant.}).

Now we have to require the condition (2.6) to be fulfilled, and this can be
easily satisfied by putting $E =0 $ in the obtained solutions, which
evidently minimizes the Hamiltonian density in (2.3) to zero value. If, for
example, one ``pumps" in some way (via interactions) the energy into particular
vacuum modes promoting their conversion to real tachyons (field excitations
satisfying relations (2.7), (2.8) at $E > 0$),
the field energy H will be increased due to an appearance of the 
kinetic energy term. 

Let us consider now a possible modification of the Lagrangian (2.1)
by adding to it an apparently Lorentz-non-invariant, but Lorentz-covariant 
term proportional to the 4-velocity $u$ of the preferred reference frame:
\begin{equation}
L = \frac{1}{2} \int d^3 {\bf x} \Big{[}\dot\Phi^2(x) -
\Big{(}\nabla\Phi(x) \Big{)}^2  + \mu^2 \Phi^2(x) +
 \lambda u^\mu \partial_\mu\Phi(x)\Big{]},
\label{eq:28}
\end{equation}
where $\lambda$ has the dimensionality of the mass squared. For the suggested
tachyon field model viability it is important that the
additional term does not change the equation of motion (2.2). 
%\begin{equation}
%\Big{(}\frac{\partial ^2}{\partial t^2} - \partial _i \partial ^i - \mu^2\Big{)} \Phi(x) = 0~, ~~~~~~~~i = 1,2,3.
%\end{equation}
Choosing  $\lambda = \mu^2$ one gets the corresponding Hamiltonian
\begin{equation}
H = \frac{1}{2} \int d^3 {\bf x} \Big{[}\dot\Phi^2(x) +
\Big{(}\nabla\Phi(x) \Big{)}^2 - \mu^2 \Phi^2 +
\frac{ \mu^2{\bf u}}{\sqrt{1-{\bf u}^2}} \nabla \Phi(x)\Big{]}.  
\label{eq:29}
\end{equation}
Thus, the additional term in the integrand of (\ref{eq:29}) shifts the tachyon 
vacuum energy boundaries depending on the value and direction of the 3-velocity
of the preferred reference frame ${\bf u}$ with respect to the tachyon source 
(illustrated by formulae (1.3), (1.4) and by fig.~1), 
which was just the aim of the introduction of this term.

\includegraphics[height=13cm,width=26cm]{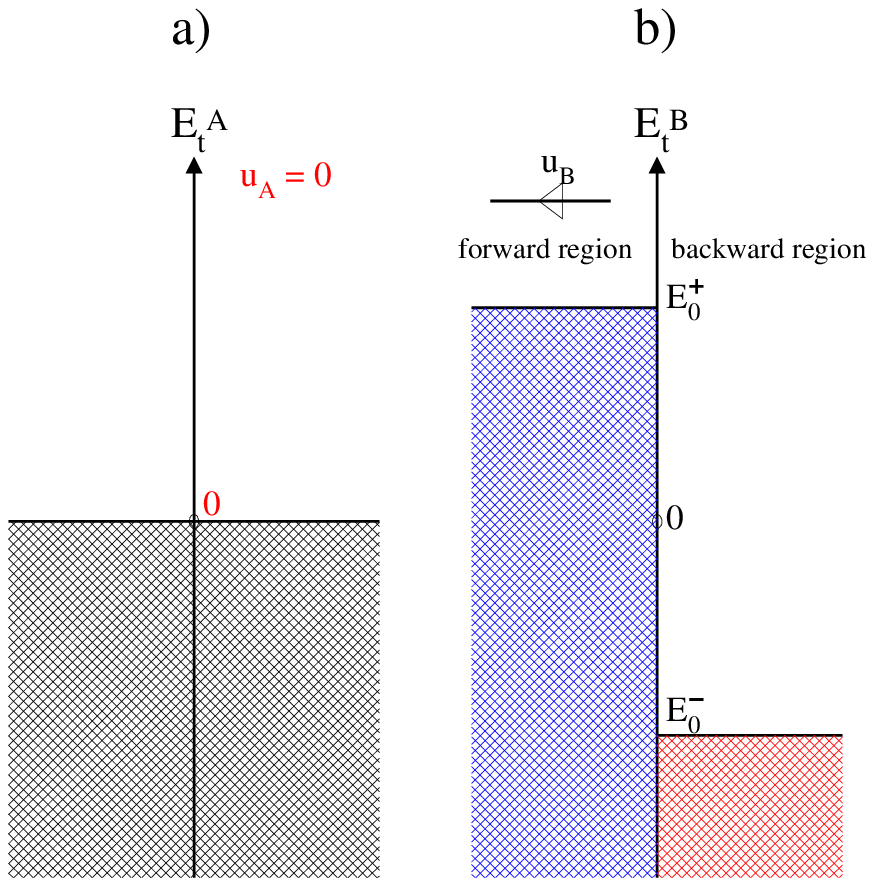}
Fig.~1. The tachyon vacuum energy boundaries as seen $a)$ from the preferred
reference frame A and $b)$ from a frame B moving with respect to the preferred 
one with 3-velocity~${\bf u}$. The direction of the preferred frame motion 
as seen from the frame B is indicated by an arrow in the top part
of $b)$. $E_{0}^{+}$ and $E_{0}^{-}$ mark the ``forward" and the ``backward"
tachyon vacuum energy boundaries in the moving frame given by (1.3) and (1.4).
The vertical axes on both figures are for tachyon energies, with the hatched
regions to be excluded domains for free tachyons.

\vskip5mm
After second quantization procedure the Hamiltonian reads (see \cite{ttheor}):
\begin{equation}
H=\int_{|{\bf k}| > \mu,\omega>{\bf ku}}
{\frac{d^3{\bf k}}{(2\pi)^3}\frac{\omega-{\bf ku}}{\sqrt{1-{\bf u}^2}}
~a^+_{{\bf k}} a_{{\bf k}}}.
\label{eq:210}
\end{equation}
Thus the Hamiltonian is bounded from below and is Hermitian. 
In the preferred reference frame
\begin{equation}
H = \int_{|{\bf k}| > \mu,\omega>0}
{\frac{d^3{\bf k}}{(2\pi)^3}~\omega ~a^+_{{\bf k}} a_{{\bf k}}}
\label{eq:211}
\end{equation}
having non-negative eigenvalues.

To conclude this section we formulate its result: the Lagrangian (\ref{eq:28}) 
differs from the Lagrangian (2.1) by a Lorentz-non-invariant term presented 
in the former; and since this additional term, written down as 
$\lambda \partial_\mu F^\mu(x)$, where $F^\mu(x) \equiv u^\mu \Phi(x)$,
is proportional to the total divergence of the 4-vector $F^\mu(x)$, the two
Lagrangians, with and without the additional term, are physically equivalent
since the term with $\partial_\mu F^\mu(x)$ does not contribute to physical
quantities, excepting those related to the tachyon vacuum.

\vskip4mm
Furthermore, since the additional term does not change the action we have,
by the principle of least action, the same tachyon equation of motion (2.2),
as mentioned above \footnote{Generally speaking, when passing from a 
Lagrangian to a Hamiltonian one should take into account that there is a term
in the Hamiltonian density proportional to a (partial) derivative of the 
Lagrangian density over $\dot\Phi$ (the momentum density conjugate to $\Phi$). 
The term with this derivative, resulting in a transition from the Lagrangian 
to the Hamiltonian, destroys the immunity of the latter from the adding of
an extra term (the gradient term in this case) to it.

The different influence of given extra terms on the Lagrangian 
and on the Hamiltonian should not surprise us: while the Lagrangian is a 
Lorentz scalar (the same is true for the action), the Hamiltonian is the 
0-component of the field 4-momentum.}. 
Therefore within our approach the Lorentz invariance can be defined as
covariantly broken and its violation appears to be restricted to
the asymptotic-tachyon-states sector only, 
as will become clear from the further considerations. 

\section{Lorentz invariance of tachyon Feynman propagator}
\setcounter{equation}{0}
\renewcommand{\theequation}{3.\arabic{equation}}
Considering a tachyon propagator 
in momentum space as an inverse of a Fourier transform of the wave 
equation (2.2) with the condition (\ref{eq:8}) imposed, we can write down, 
for example, the Feynman propagator as
\begin{equation}
\tilde D_F(k) = \frac{i~\theta(|\bf{k}| - \mu)}{k^2 + \mu^2 + i\epsilon}  
\end{equation}
to be used in Feynman diagrams describing tachyon interactions, of course, 
only within our toy model of scalar tachyons. In the configuration space
\begin{equation}
D_F(x-y) = \int_{|\bf{k}| \geq \mu} \frac{d^4 k} {(2\pi)^4}~ 
\frac{i~\exp{[-ik(x-y)]}} {k^2+\mu^2+i\epsilon}. 
\end{equation}
One can see that the tachyon Feynman propagator defined by formula (3.2) 
is explicitly Lorentz-invariant since all ingredients in this formula, 
including integration limits, are Lorentz-invariant (in particular, if the
integration limit $|\bf{k}| \geq \mu$ holds in some inertial frame it holds 
in any such frame \footnote{The proof of this statement is simple. If we 
consider the condition $|\bf{k}| \geq \mu$ in combination with the tachyon 
mass-shell hyperboloid $k^2 + \mu^2 = 0$ at the tachyon energy $k^0 = 0$, 
i.e. acting along the $|\bf{k}|$ axis, its validity extends over the range from 
$|\bf{k}|= \mu$ to $|\bf{k}| = \infty$. Since the $k^2$ is a Lorentz invariant 
this means that the above condition is valid in the whole external volume 
outside the tachyon mass shell.}).

Let us note that, as a consequence, the same invariance holds also for 
virtual tachyons and tachyon loops appearing in Feynman diagrams of reactions 
containing only ordinary particles in the initial and final states; thus
no Lorentz-violating effects induced by virtual tachyons can appear in such
reactions. Representations of the tachyon Feynman propagator (3.2) 
in configuration space are given in the next section.

It is not a problem (see Section~5) to obtain (3.2) as a time-ordered product 
of the tachyon field operators (1.5), taken at points $x$ and $y$ and averaged 
over the tachyonic vacuum, i.e. as an amplitude of the tachyon transition 
from $x$ to $y$ or {\em vice versa}: 
\begin{equation}
 D_F(x-y) \equiv \langle 0|T \Phi(x) \Phi(y)|0\rangle =
\cases{D(x-y)~~~& if~$x^0  >  y^0$ \cr
       D(y-x)~~~& if $x^0  <  y^0$ \cr },
\label{eq:33}
\end{equation}
where $D(x-y)$ is a correlation function of the tachyon field operators 
$\Phi(x), \Phi(y)$:
\begin{equation}
D(x-y) \equiv  \langle 0|\Phi(x) \Phi(y)|0\rangle = 
\int_{|\bf{k}| \geq \mu} \frac{d^3 {\bf k}} {(2\pi)^3}~
\frac{\exp{[-ik(x-y)]}}{2 \omega}
\end{equation}

According to the Lehman-Symanzik-Zimmerman reduction formula, in the
computation of the $S$-matrix elements using the Feynman diagrams the Feynman
propagators should be attached to the internal lines of the diagrams only.
Therefore the Feynman rules for the construction of Feynman diagrams 
which include tachyons appear to remain standard, with a minor exception: 
to each external tachyon leg (asymptotic tachyon state) the causal $\Theta$ 
function $\Theta(ku)$ should be attached to ensure a cut of the phase space 
of reactions with such states as described in \cite{pvmich}.  

We observe thus that the operation of the time ordering of field operators
in (\ref{eq:33}) realises, in general, the same function 
as causal $\Theta$-terms in the field operators (1.5). 
Therefore we can interpret heuristically the obtained results as follows.   

The standard $i\epsilon$ prescription, aimed at the definition of the 
integration contour on the complex plane of $k^0$, allows virtual tachyons
(as well as virtual ordinary particles) to avoid causal restrictions, which are
otherwise (i.e. in the case of real faster-than-light particles) imposed by the
causal $\Theta$ function on the propagation of free tachyons. In other words, 
the $i\epsilon$ prescription, ensuring the time ordering of the tachyon field 
operators which contain the $\Theta$ functions apparently breaking the Lorentz 
invariance, makes the virtual tachyons insensitive to the existence of the
preferred reference frame (which indeed is intuitively obvious: virtual
tachyons cannot be used as signal carriers), and this
results in the Lorentz invariance of the tachyon Feynman propagator, 
similarly to the Feynman propagators of ordinary particles. 

The Lorentz invariance of the tachyon Feynman propagator, obtained in our model
with the covariantly broken Lorentz invariance, means, together with the
assumed Lorentz invariance of the Feynman propagators of all other particles, 
that the speed of light remains a unique, universal velocity constant which 
limits particle velocities on both sides of the light barrier, bounding the 
maximum attainable velocities of ordinary particles and restricting the tachyon 
velocities from below. In particular, an explicit breaking of the Lorentz 
symmetry by adding to the Lagrangian the Lorentz-violating terms which affect 
the particle propagators, suggested by authors of the Standard Model Extension 
(SME) \cite{kostel} (see also \cite{colgla1,colgla2}), which lead
to individual maximum attainable velocity for each fundamental field, 
differing from the velocity of light, is not relevant to our approach.
For the same reason the strong restrictions on multiple Lorentz-violating
coefficients compiled in the ``Data Tables for Lorentz and CPT violation"
\cite{kostel2} (see also \cite{mattingly}) are not applicable to our 
considerations.

\section{The Feynman propagator for scalar tachyons in the configuration space}
\setcounter{equation}{0}
\renewcommand{\theequation}{4.\arabic{equation}} 
Let us obtain $D_F(x-y)$ explicitly:
\begin{eqnarray}
D_F(x-y) = \int_{|\bf{k}| \geq \mu} \frac{d^4 k} {(2\pi)^4}~ 
\frac{i~\exp{[-ik(x-y)]}} {k^2+\mu^2+i\epsilon}
\nonumber \\
= \int_{|\bf{k}| \geq \mu} \frac{d^3 {\bf k}~d k_0} {(2\pi)^4}~ 
\frac{i~\exp{[-i k_0 \Delta t + i{\bf k(x-y)}]}} {k^2+\mu^2+i\epsilon}
\nonumber \\
= \int_{|\bf{k}| \geq \mu} \frac{d^3 {\bf k}}{(2\pi)^3 2 \omega} 
~\exp[-i\omega |\Delta t| + i{\bf k(x-y)}], 
\end{eqnarray}
where $\Delta t = x^0 - y^0$, and we have used the integral representation
\begin{equation}
\exp{(-i\omega |\Delta t|)} = \frac{i\omega}{\pi}
\int_{-\infty} ^{+\infty} d k_0~
\frac{\exp (-i k_0 \Delta t)}{k_0^2 - \omega^2 + i\epsilon},
~~~~~~~~~~~~\epsilon \rightarrow 0^+ . 
\end{equation}
Integration of (4.1) over the angles of ${\bf k}$ gives
\begin{equation}
D_F(x-y) = \frac{1} {4 \pi^2} \int_{\mu} ^{\infty} dk \frac{{\bf k}^2}{\omega}~
\frac{\sin(|{\bf k}||{\bf x-y}|)} {|{\bf k}||{\bf x-y}|} 
~\exp(-i\omega |\Delta t|).
\end{equation}
With the definition of the interval
\begin{equation}
s \equiv \Delta t^2 - {\bf (x-y)}^2 = \Delta t^2 -  r^2,
\end{equation}
where $r \equiv |{\bf (x-y)}|$, 
we can investigate the behaviour of the $D_F$ outside and inside the light 
cone. For spacelike intervals, $s < 0$, we can put $\Delta t = 0$ to obtain
\begin{equation}
D_F(r) =  \frac{1}{4\pi^2 r} \int_{\mu}^{\infty} \frac{|{\bf k}| dk}
{\sqrt{{\bf k}^2 - \mu^2}}~\sin|{\bf k}| r =-\frac{ \mu}{8\pi r}~Y_1(\mu r), 
\end{equation} 
where $Y_1$ is the Bessel function of the second kind; it represents an
outgoing wave for large $r$ indicating the allowance of tachyon asymptotic
states. To compare: for an ordinary scalar particle with 
the mass $m$ the corresponding Feynman propagator
\begin{equation}
D_F^{ord}(r) =  \frac{ m}{4\pi^2 r}~K_1(mr),
\end{equation}
where $K_1$ is the modified Bessel function of the second kind \cite{huang}; 
it damps exponentially for large $r$, the characteristic damping length being
the particle Compton length $\lambda = 1/m$. 

For timelike intervals, $s > 0$, we can put $|{\bf x}-{\bf y}| \rightarrow 0$ 
in (4.3), i.e. $r = 0$ in (4.4): 
\begin{equation}
D_F(|\Delta t|) = \frac{1}{4\pi^2} \int_{\mu}^{\infty} \frac{{\bf k}^2 dk}
{\sqrt{{\bf k}^2 - \mu^2}} \exp{(-i\sqrt{{\bf k}^2 - \mu^2}~|\Delta t|)} 
= \frac{ \mu}{4\pi^2 |\Delta t|}~K_1(\mu |\Delta t|),
\end{equation}
i.e. it damps exponentially for large $|\Delta t|$, with the characteristic 
damping time being the tachyon Compton length $\lambda = 1/\mu$. 
For an ordinary scalar particle
\begin{equation}
D_F^{ord}(|\Delta t|) = \frac{im}{8\pi |\Delta t|}~H_1 ^{(1)} (m |\Delta t|),
\end{equation}
where $ H_1 ^{(1)}$ is the Hankel function of the first kind 
which represents an outgoing wave for large $|\Delta t|$ \cite{huang}. 

\section{Lorentz-invariant and Lorentz-non-invariant 
two-point functions of scalar tachyon fields}        
\setcounter{equation}{0}
\renewcommand{\theequation}{5.\arabic{equation}}
The correlation function $D(x-y)$ can be represented as
\begin{eqnarray}
D(x-y) \equiv \langle 0|\Phi(x) \Phi(y)|0 \rangle =
\Big{\langle}
\int 
\frac{d^4k}{(2\pi)^3}~\exp[-ik(x-y)]~\delta(k^2+\mu^2)~\Theta(ku)
~\Big{\rangle}_{Evac}
\nonumber \\
=
\Big{\langle}
\int_{|\bf{k}| \geq \mu,\omega \geq {\bf ku}} 
\frac{d^3{\bf k}} {(2\pi)^3}~\frac{\exp{[-i\omega \Delta t + i{\bf k(x-y)}]}}
{2\omega}
~\Big{\rangle}_{Evac} 
\end{eqnarray}
where $\Delta t = x^0 -y^0$ 
and the angular brackets $\langle~\rangle_{Evac}$,
surrounding the integrals in (5.1), denote the averaging over the tachyon 
vacuum energy boundaries. Such an averaging is a necessary action 
since the tachyon vacuum energy boundaries are not, in general (in the frames 
moving with respect to the preferred one), rotationally invariant.   
   
Generally speaking, the calculation of tachyon vacuum expectation 
values of any combination of tachyon operators requires such an averaging 
as distinct to the calculation of analogous vacuum expectation values in the 
case of ordinary particles, when such calculations result in Lorentz-invariant 
c-number functions due to Lorentz-invariance of the ordinary particle vacuum. 
In our case (with tachyons) the expression inside the vacuum brackets 
$\langle~\rangle_{Evac}$ in (5.1) is a Lorentz-non-invariant c-number function 
due to the Lorentz-non-invariant (though Lorentz-covariant) integration limits,
$\omega \geq {\bf ku}$ in (5.1), coming from the causal 
$\Theta$-function (these limits are illustrated, in particular, by formulae 
(1.3), (1.4) and by fig.~1). Note, an observer in a frame moving with respect 
to the preferred one also can detect that the energy boundaries of the tachyon 
vacuum are different in the forward and backward hemispheres of his motion 
\footnote{An attempt to detect these boundaries was undertaken 
in \cite{pvmich}.}. 

Fortunately, 
the averaging of the expression inside the vacuum brackets in (5.1) over 
the tachyon vacuum energy boundaries
contracts the above Lorentz-non-invariance since these boundaries are governed 
by formula (1.2), the same one which imposes those integration limits. 
This occurs owing to the fact that the boundaries are symmetric with respect to
the zero energy level in the preferred reference frame, 
depending on the direction of the observer's motion.
The statement about Lorentz-invariance of the correlation function $D(x-y)$ 
can be proved as follows.

First of all, 
we note that the averaging over the tachyonic vacuum energy boundaries can be 
done for each individual direction of ${\bf k}$, i.e. for fixed values of the 
angles $\theta$ and $\phi$, the polar and azimuthal angles with respect to the
directions of ${\bf k}$ and ${\bf x - y}$. Changing the first integration
over $d^4k$ in (5.1) from $d\omega$ to $d|{\bf k}|$ we can rewrite it as
\begin{eqnarray}
{D(x-y) =                                                  
\int \Bigg{[}\frac{d \cos{\theta} d\phi} {2(2\pi)^3}}~~~~~~~~~~~~~~~~~~~~~~~~~~~~~~~~~~~~~~~~~~~~~~~~~~~~~~~~~~~~~~~~~~~~~~~~~~~~~~
\nonumber \\
\times 
\Big{\langle} 
\int_{\omega \geq \sqrt{\omega^2+\mu^2} |{\bf u}| \cos{\psi}} 
d\omega \sqrt{\omega^2 + \mu^2} 
\exp{(-i\omega \Delta t+i\sqrt{\omega^2+\mu^2}|{\bf x-y}|\cos\theta)}~
\Big{\rangle}_{Evac}
\Bigg{]}~
\end{eqnarray}
Here $\psi$ is the angle between the directions of ${\bf k}$ and ${\bf u}$. 
Obviously, the expression enclosed by the vacuum brackets 
in (5.2) can be written as a sum
\begin{equation} 
\frac{1}{2} \Big{(} I_{FW}({\bf x-y,u}) + I_{BW}({\bf x-y,u})\Big{)}, 
\end{equation}
where
\begin{equation}
I({\bf x-y,u}) = 
\int_{\omega \geq \sqrt{\omega^2+\mu^2} |{\bf u}| \cos{\psi}} 
d\omega \sqrt{\omega^2 + \mu^2} 
\exp{(-i\omega \Delta t + i\sqrt{\omega^2+\mu^2}|{\bf x-y}|\cos\theta)},
\end{equation}
and the subscripts $FW$ and $BW$ define the corresponding integrals in the 
forward and backward hemispheres of the vector ${\bf u}$, where $\cos{\psi} >0$
and  $\cos{\psi} < 0$, respectively.

Each of the hemisphere integrals $I({\bf x-y,u})$ can be written down 
as a combination of two terms, one of which does not depend on the vector 
${\bf u}$, and another which does. So, 
\begin{eqnarray}
I_{FW}({\bf x-y,u}) 
= \int_{0}^{\infty} d\omega \sqrt{\omega^2 + \mu^2} 
\exp{(-i\omega \Delta t + i \sqrt{\omega^2 + \mu^2}|{\bf x-y}|\cos\theta)} 
\nonumber\\
- \int_{0}^{E_0^+(|{\bf u}|,\psi)} d\omega \sqrt{\omega^2 + \mu^2} 
\exp{(-i\omega \Delta t + i\sqrt{\omega^2 +\mu^2}|{\bf x-y}|\cos\theta),}
\end{eqnarray}
where
\begin{equation} 
E_0^+(|{\bf u}|,\psi)= \frac{\mu |{\bf u}| \cos\psi}
{\sqrt{1 - {\bf u}^2 }},
~~~~~~\cos\psi > 0. 
\end{equation}

Analogously,
\begin{eqnarray}
I_{BW}({\bf x-y,u})  
= \int_{0}^{\infty} d\omega \sqrt{\omega^2 + \mu^2}
\exp{(-i\omega \Delta t + i \sqrt{\omega^2 + \mu^2}|{\bf x-y}|\cos\theta)} 
\nonumber\\
+\int_{E_0^-(|{\bf u}|,\psi)}^{0} d\omega \sqrt{\omega^2 + \mu^2}  
\exp{(-i\omega \Delta t + i\sqrt{\omega^2 +\mu^2}|{\bf x-y}|\cos\theta)},
\end{eqnarray}
where
\begin{equation} 
E_0^-(|{\bf u}|,\psi)= \frac{\mu |{\bf u}| \cos\psi}
{\sqrt{1 - {\bf u}^2}},
~~~~~~\cos\psi < 0, 
\end{equation}

As follows from (5.6), (5.8), $|E_0^-(|{\bf u}|,\psi)|=E_0^+(|{\bf u|},\psi)$.
Therefore\footnote{Probably, it is worth noting
that the exponential terms in (5.9) below are identical 
even though the integration domains over the variable $\omega$ in them are 
quite different: this is due to the fact that exponential indices, containing
the $\omega$, are scalar products of the 4-vectors $k$ and $x-y$ (multiplied
by $i$), i.e. they are Lorentz scalars by definition.}       
\begin{eqnarray}
\int_{0}^{E_0^+(|{\bf u}|,\psi)} d\omega \sqrt{\omega^2 + \mu^2} 
\exp{(-i\omega \Delta t + i \sqrt{\omega^2 + \mu^2}|{\bf x-y}|\cos\theta)} 
\nonumber\\
=\int_{E_0^-(|{\bf u}|,\psi)}^{0} 
d\omega \sqrt{\omega^2 + \mu^2} 
\exp{(-i\omega \Delta t + i\sqrt{\omega^2 +\mu^2}|{\bf x-y}|\cos\theta)}.
\end{eqnarray}
As a result
\begin{eqnarray}
\frac{1}{2}\Big{(}I_{FW}({\bf x-y,u}) + 
I_{BW}({\bf x-y,u})\Big{)}~~~~~~~~~~~~~~~~~~~~~~~
\nonumber\\
=\int_{0}^{\infty} d\omega \sqrt{\omega^2 + \mu^2} 
\exp{(-i\omega \Delta t + i\sqrt{\omega^2 +\mu^2}|{\bf x-y}|\cos\theta)}.
\end{eqnarray}

Collecting all the ingredients of (5.2) and reverting back 
from the integration over $d\cos{\theta} d\phi ~\omega d\omega$ 
to the integration over $d^3{\bf k}$ we obtain finally 
\begin{equation}
D(x-y) = \int_{|\bf{k}| \geq \mu} \frac{d^3 {\bf k}} {(2\pi)^3}~ 
\frac{\exp{[-i\omega \Delta t + i{\bf k(x-y)}]}}{2 \omega},
\end{equation}  
which does not depend on ${\bf u}$, i.e. it is manifestly Lorentz-invariant,
as has been stated above.

This function can be used to construct the Feynman propagator
\begin{equation}
D_F(x-y) = \cases{D(x-y)~& if~$x^0  >  y^0$ \cr
                  D(y-x)~& if $x^0  <  y^0$ \cr }
= \int_{|\bf{k}| \geq \mu} \frac{d^3 {\bf k}} {(2\pi)^3}~ 
\frac{\exp{[-i\omega |\Delta t| + i{\bf k(x-y)}]}}{2 \omega}
\label{eq:512}
\end{equation}
in which we have changed the integration variable from ${\bf k}$ to ${-\bf k}$
when replacing $\Delta t$ by $|\Delta t|$ in the case of $x^0  <  y^0$. 
The resulting expression in (\ref{eq:512}) coincides with the last line of 
(4.1) which proves (\ref{eq:33}). 

The commutator of scalar tachyon fields reads \cite{ttheor}
\begin{equation}
\Delta(x-y) \equiv[\Phi(x),\Phi(y)] = \int{\frac{d^4k}{(2\pi)^3}~\Big{\{}}{\exp[-ik(x-y)] - 
\exp[(ik(x-y)]\Big{\}}~\delta(k^2+\mu^2)~\Theta(ku)}
\end{equation}
which is not automatically zero at $(x-y)^2 < 0$ as distinct to the field
commutators of ordinary particles.
Let us consider it in the preferred reference frame:
\begin{eqnarray}
[\Phi(x),\Phi(y)] = \frac{1}{(2\pi)^3}
\int_{|{\bf k}| \geq \mu,\omega \geq 0} \frac{d^3{\bf k}}{2\omega}~
\Big{\{}\exp{[-i \omega \Delta t + i{\bf k(x-y)]}}
\nonumber \\
-\exp{[i\omega \Delta t - i{\bf k(x-y)]}}\Big{\}}
\end{eqnarray}
If $\Delta t \neq 0$ the commutator does not vanish (excepting the case of
$\omega = 0$ corresponding to the exchange of vacuum tachyons).
However, the commutator $\Delta(x-y)$ vanishes at $\Delta t = 0$
in the preferred reference frame, which is obvious from (5.14); the same 
is true for the correspondingly Lorentz-shifted $x-y$ in boosted frames
(due to the covariance of the expression (5.13)).
 
Though being Lorentz-covariant, the commutator (5.13) is not Lorentz-invariant
since it depends on the value of $|{\bf u}|$ and
on the angle between the directions of ${\bf u}$ and ${\bf x - y}$
in the frames moving with respect to the preferred one. In other words, 
though all terms in the integrand of (5.13), except $\Theta(ku)$, are Lorentz
invariant, the integration limits in this expression, imposed by the causal
$\Theta$-function, are rotationally invariant 
in the preferred reference frame only, as can be seen from (5.14).
Just this general rotational non-invariance of the integration limits results 
in the Lorentz-non-invariance of the commutator (5.13).  
This means that the amplitude of propagation of
a tachyon from $x$ to $y$ is not equal to the amplitude of propagation of
the same tachyon from $y$ to $x$ \footnote{In the case of a complex tachyon 
field the propagation of the tachyon from $y$ to $x$ would be replaced by the
propagation of an antitachyon in that direction.}, the latter being
the complex conjugate of the former. 

However, being averaged over the tachyonic vacuum, the commutator
\begin{equation}
\langle 0|[\Phi(x),\Phi(y)]|0 \rangle
\end{equation} 
becomes Lorentz-invariant since the rotational non-invariance mentioned in the 
previous paragraph is cancelled by the averaging. Simultaneously, this leads
to the vanishing of the commutator (5.15) at spacelike separations
of the $x$ and $y$ positions: 
\begin{equation}
\langle 0|\Delta(x-y)|0 \rangle \equiv \langle 0|[\Phi(x),\Phi(y)]|0 \rangle=
D(x-y) - D(y-x) = 0~~~~\Big{(}(x-y)^2 < 0\Big{)}.  
\end{equation}
Expression (5.16) can be considered as a limiting case of a causal loop 
construction with the use of tachyons. Its vanishing means a principal 
impossibility of such a construction expressed in terms of quantum field 
theory, i.e. (5.16) is {\em the micro-causality condition of a tachyon theory.}

Finally, the equal-time commutator of the scalar tachyon field with its 
canonical conjugate $\dot\Phi$ does not vanish \cite{ttheor}:
\begin{equation}
[\Phi(x),\dot\Phi(y)]_{x^0=y^0} = i~\bar\delta^3 {\bf(x-y}), 
\end{equation}
where $\bar\delta^3 ({\bf x-y})$ is a $truncated$ delta function which acts
like the standard delta function with respect to those functions whose Fourier
transforms vanish at $|{\bf k}| < \mu, ~\omega < ({\bf ku})$ (acting similarly 
to the truncated delta functions introduced in \cite{fein,arsud}); with this 
specification (5.17) corresponds to the analogous commutator for ordinary 
scalar particle fields containing the $standard$ delta function.

\section{Summary} 
The overall approach to tachyon field models emerging from the consideration 
presented in this note possesses the following attractive features:    
\begin{itemize}
\item [1.] Its main concept is based on the experimentally proved phenomenon
\cite{pdg}: the existence of a preferred reference frame which is the comoving 
frame of the relativistic cosmology \cite{ll1}.
\item [2.] The introduction of this concept into the tachyon theory can be
done in a covariant way which leads to the result that the Lorentz 
symmetry of the tachyon theory appears to be covariantly broken.
\item [3.] The covariant breaking of the Lorentz symmetry allows one 
to avoid introducing ``by hand" Lorentz-violating terms into Lagrangians 
as it is suggested by authors of the Standard Model Extension (SME) 
\cite{kostel,colgla1,colgla2},
which would lead to the Lorentz-non-invariant tachyon propagators 
making tachyons subject to strict restrictions imposed on the Lorentz violation
by a variety of experiments (see \cite{kostel2,mattingly}). 
\item [4.] Faster-than-light velocities of particles and signals
appear to be allowed in a wide range of $c < v < \infty$ as distinct to 
the standard models of the Lorentz invariance violation (i.e. SME) in which 
only tiny positive deviations from the velocity of light are permitted 
for particles having energies below the Plank mass scale.
%\cite{kostel,colgla1,colgla2}.  
\item [5.] So formulated the tachyon hypothesis results in the possibility of 
the existence of a new world of elementary particles residing beyond the light 
barrier as distinct to the SME 
in which the Lorentz violating particles are assumed to be (some of the) 
already known particles, {\em e.g.} neutrinos, electrons, muons, pions, etc.
\item [6.] The velocity of light remains an invariant quantity to be a barrier
between the tachyon world of particles and that of subluminal particles
which can be considered as a proof of a fundamental character of this
velocity. 
\end{itemize}

\section{Conclusion}
A modification of a scalar tachyon field Lagrangian by adding to it a 
Lorentz-non-invariant, but Lorentz-covariant term is suggested, aimed at
the conservation of causality, with the Lorentz invariance 
of the tachyon Feynman propagator being conserved.
It is shown that the standard $i\epsilon$ prescription in the Feynman 
propagator realises the same function over virtual particles (having 
$k^2 \geq 0$ as well as $k^2 < 0$) as the causal $\Theta$ function in the 
quantum field operators of free tachyons (asymptotic tachyon states).  
The result is a possibility to have a tachyon model free of causal paradoxes,
tachyon vacuum instability, and Lorentz-violating radiative corrections 
coming from virtual tachyons to processes involving only ordinary particles,  
restricting Lorentz-violating effects to the free-tachyon sector only.

\section*{Acknowledgements}
I express my gratitude to Profs.~Boreskov~K.~G., Dzheparov~F.~S., 
Kancheli~O.~V. and Sibiryakov~S.~M. for stimulating discussions, 
and to Dr.~French~B.~R. for the critical reading of the manuscript.
After this article has been completed I have obtained the information 
from Dr.~Giudici~G.~F., to whom I am indebted very much, about the paper 
\cite{radzik}, in which a stable, renormalizable, scalar tachyonic quantum 
field theory with chronology protection is suggested, developed with an
approach quite different to that presented here and in \cite{ttheor}.
 
%This research did not receive any specific grant from funding agencies in the 
%public, commercial, or not-for-profit sectors.
 
%\newpage

\end{document}